\input{aipcheck}

\documentclass[
    ,final            
]
  {aipproc}
\usepackage{epsf,epsfig,amssymb,latexsym,amsmath,amsthm,psfrag}
\layoutstyle{6x9}

\newcommand{\be}{\begin{equation}}
\newcommand{\ee}{\end{equation}}
\newcommand{\ba}{\begin{eqnarray}}
\newcommand{\ea}{\end{eqnarray}}

\newcommand{\om}{\Omega}
\newcommand{\nn}{\nonumber}

\begin{document}

\title{S-wave $\gamma\gamma\to \pi\pi$ and $f_0(980)\to\pi\pi$}


\author{J. A. Oller}{
  address={Departamento de F\'{\i}sica, Universidad de Murcia }
}

\author{L. Roca}{
  address={Departamento de F\'{\i}sica, Universidad de Murcia }
}

\author{C. Schat}{
address={CONICET and Departamento de F\'{\i}sica, FCEyN, Universidad de Buenos Aires, \\
Ciudad Universitaria, Pab. 1, (1428) Buenos Aires, Argentina}
}

\begin{abstract}
We report on a dispersion relation for the  
 $\gamma\gamma\to (\pi\pi)_I$ S-wave in isospin $I$ emphasizing
 the low energy region. 
The $f_0(980)$ signal that emerges in $\gamma\gamma\to \pi\pi$ is 
also discussed. Our results could be used to distinguish between different 
$\pi\pi$ isoscalar S-wave parameterizations. 
 We also calculate the width of the $\sigma$ resonance to $\gamma\gamma$ and 
 obtain the 
 value $\Gamma(\sigma\to\gamma\gamma)=(1.68\pm 0.15)~$KeV. Finally, we elaborate 
 on the size of the $f_0(980)$ coupling to $\pi\pi$ and show that its smallness compared 
 to the $K\bar{K}$ one is not related to the OZI rule.
\end{abstract}

\maketitle


\section{Introduction}

The study of the reaction $\gamma \gamma \to \pi^0\pi^0$ offers the
possibility of having  two neutral pions in the final state as a
two body hadronic system.  Because  $\gamma\gamma\to \pi^0\pi^0$ has no  Born term
 the final state interactions 
dominate this reaction.  In this respect,
one can think  of the interest of precise data on this process to
test present low energy parameterizations of the $\pi\pi$ isospin ($I$) 0 S-wave, 
 to further study the nature of the $\sigma$ resonance 
 as well as that of the $f_0(980)$, and having another way
to constraint their pole positions \cite{caprini,alba}. The $\sigma$ coupling to $\gamma\gamma$ is discussed 
and we obtain  $\Gamma(\sigma\to \gamma\gamma)=(1.68\pm 0.15)~$KeV.  
Finally, we show that the suppression of the $f_0(980)$ coupling to $\pi\pi$ in comparison with
$K\bar{K}$ is not due to the  OZI rule, as one could 
be tempted to think \cite{polosa}.

\section{$\gamma\gamma\to \pi\pi$}

For this and the next section we
 report on the results of refs.\cite{orsletter,orlast}, where a more detailed 
account can be found. 
Let us consider the S-wave amplitude $\gamma\gamma\to(\pi\pi)_I$,
$F_I(s)$, where the two pions have definite $I=0$ or 2. 
 The function $F_I(s)$ on the complex $s-$plane is analytic except for two cuts along the real 
 $s-$axis, 
 the unitarity one for $s\geq 4m_\pi^2$ and the left hand cut for $s\leq 0$, with 
 $m_\pi$ the pion mass. Let us 
denote by $L_I(s)$ the complete left hand cut contribution to $F_I(s)$. Then, the
 function $F_I(s)-L_I(s)$, by construction, has only right hand cut.  
 Let $\phi_I(s)$ be the phase of $F_I(s)$ modulo $\pi$, chosen in such a way that 
 $\phi_I(s)$ is {\it continuous} and $\phi_I(4m_\pi^2)=0$. 
 For the exotic $I=2$ S-wave one can invoke Watson's final state theorem\footnote{This theorem implies 
that the phase of $F_I(s)$ when there is no inelasticity is the same, modulo $\pi$,
 as the one of the isospin $I$ S-wave $\pi\pi$ elastic strong amplitude.} so that 
$\phi_2(s)=\delta_\pi(s)_2$.  
For $I=0$ the same theorem guarantees that $\phi_0(s)=\delta_\pi(s)_0$ for $s\leq 4m_K^2$,  
where we denote by $\delta_\pi(s)_I$ the isospin $I$ S-wave $\pi\pi$ phase shifts.  
Here one neglects the inelasticity due to the $4\pi$ and $6\pi$
states below the two kaon threshold \cite{hyams}.
 Above the two kaon threshold $s_K=4m_K^2$, 
  the phase function $\phi_0(s)$ cannot 
be fixed {\it a priori} due 
the onset of inelasticity. 
However, as remarked in refs.\cite{y04,or07}, inelasticity is again small 
for $\sqrt{s}\gtrsim 1.1$~GeV  \cite{hyams}, and   one can then apply 
 approximately Watson's final state 
 theorem which implies that 
  $\phi_0(s)\simeq \delta^{(+)}(s)$ modulo $\pi$. Here  $\delta^{(+)}(s)$ is the eigenphase 
  of the $\pi\pi$, $K\bar{K}$ $I=0$ S-wave S-matrix 
 such that it is continuous and $\delta^{(+)}(s_K)=\delta_\pi(s_K)_0$. In refs.\cite{y05,or07} it is
 shown that $\delta^{(+)}(s)\simeq \delta_\pi(s)_0$ or $\delta_\pi(s)_0-\pi$, depending on whether 
 $\delta_\pi(s_K)_0\geq \pi$ or $<\pi$, respectively. In order to fix the 
integer factor in front of $\pi$ in the relation 
$\phi_0(s)\simeq \delta^{(+)}(s)$ modulo $\pi$, it is necessary to follow the possible 
 trajectories of  $\phi_0(s)$ 
 in the {\it narrow} region $1\lesssim \sqrt{s}\lesssim 1.1$~GeV. 
 The remarkable physical effects happening there 
 are the appearance of the 
 $f_0(980)$ resonance on top of the $K\bar{K}$ threshold and the cusp effect of the latter 
 that induces a discontinuity  at $s_K$ in the derivative  of observables. 
 Between 1.05 to 1.1~GeV there are no further narrow structures and observables evolve smoothly. 
 Approximately half of the region between 
0.95 and 1.05~GeV is elastic and $\phi_0(s)= \delta_\pi(s)_0$ (Watson's theorem), so that it raises 
 rapidly. Above $2 m_K\simeq 1$~GeV and up to 1.05~GeV the function $\phi_0(s)$ can keep increasing with 
 energy, like $\delta_\pi(s)_0$. 
  The other possibility is a change  of sign in the slope at $s_K$ due to the 
$K\bar{K}$ cusp effect such 
 that $\phi_0(s)$  starts a rapid decrease in 
 energy.
  Above $\sqrt{s}=1.05$~GeV,  $\phi_0(s)$ matches smoothly with the 
 behaviour for $\sqrt{s}\gtrsim 1.1$~GeV, which is  constraint 
 by Watson's final state theorem. 
   As a result, for $\sqrt{s}\gtrsim 1$~GeV 
  {\it either}  $\phi_0(s)\simeq \delta_\pi(s)_0$ {\it or}
   $\phi_0(s)\simeq \delta_\pi(s)-\pi$, corresponding
 to an increasing or decreasing $\phi_0(s)$  above $s_K$, in order.  
 
 Let us define the switch $z$ to characterize the behaviour of $\phi_0(s)$
  for $s>s_K$, and close to $s_K$,   such that
  $z=+1$ if $\phi_0(s)$ rises with energy and $z=-1$ if it decreases.
 Let $s_1$ be the value of $s$ at which $\phi_0(s_1)=\pi$. 
 Following ref.\cite{or07} we introduce the  
  Omn\`es functions,
\ba
\om_0(s)&=&\left(1-\theta(z)\frac{s}{s_1}\right)
\exp\left[ \frac{s}{\pi}\int_{4m_\pi^2}^\infty \frac{\phi_0(s')}{s'(s'-s)} ds'\right]~,\nn\\
 \om_2(s)&=&\exp\left[ \frac{s}{\pi}\int_{4m_\pi^2}^\infty \frac{\phi_2(s')}{s'(s'-s)} ds'\right]~.
\label{om0}
\ea
with $\theta(z)=1$ for $z=+1$ and 0 for $z=-1$. 
 Given the 
 definition of the phase function $\phi_I(s)$  the function 
 $F_I(s)/\om_I(s)$ has no right hand cut. Next, we perform  a twice subtracted dispersion
  relation for $(F_0(s)-L_0(s))/\om_0(s)$
\be
F_0(s)=L_0(s)+c_0 s\om_0(s)+
\frac{s^2}{\pi}\om_0(s)\int_{4m_\pi^2}^\infty
\frac{L_0(s')\sin\bar\phi_0(s')}{{s'}^2(s'-s)|\om_0(s')|}ds'
+\theta(z)\frac{\omega_0(s)}{\omega_0(s_1)}\frac{s^2}{s_1^2}(F_0(s_1)-L_0(s_1))~,
\label{f0f}
\ee
where 
$\omega_0(s)=\exp\left[ \frac{s}{\pi}\int_{4m_\pi^2}^\infty \frac{\phi_0(s')}{s'(s'-s)} ds'\right]$ 
\cite{penmorgan}. 
In the previous equation we introduce $\bar\phi_0(s)$ that is defined 
 as the phase of $\om_0(s)$.
 Proceeding similarly for $I=2$ one has
 \be
F_2(s)=L_2(s)+c_I s\,\om_2(s)+
\frac{s^2}{\pi}\om_2(s)\int_{4m_\pi^2}^\infty
\frac{L_2(s')\sin\phi_2(s')}{{s'}^2(s'-s)|\om_2(s')|}ds'~.
\label{f2f}
\ee

It is worth mentioning 
 that eq.(\ref{f0f}) for $I=0$ and $z=+1$ is equivalent to perform a three times subtracted
dispersion relation for $(F_0(s)-L_0(s))/\omega_0(s)$,
 \be
F_0(s)=L_0(s)+c_0 s\, w_0(s)+d_0 s^2 w_0(s)+\frac{s^3 w_0(s)}{\pi}\int_{4m_\pi^2}^\infty \frac{L_0(s')
\sin\phi_0(s')}{s'^3(s'-s)|\omega_0(s')|}ds'~.
\label{f0fthree}
 \ee

Let us denote by $F_N(s)$ the  S-wave $\gamma\gamma\to \pi^0\pi^0$ amplitude and by $F_C(s)$ the
$\gamma\gamma\to\pi^+\pi^-$ one. 
The relation between $F_0$, $F_2$  and $F_N(s)$, $F_C(s)$ is 
\be
F_N(s) = -\frac{1}{\sqrt3} F_0 + \sqrt{\frac23} F_2~, ~
F_C(s) = -\frac{1}{\sqrt3} F_0 - \sqrt{\frac16} F_2 ~.
\ee

We are still left with the unknown subtraction 
constants $c_0$, $c_2$ for $I=0$ and $2$, respectively, and 
$F_0(s_1)-L_0(s_1)$ for $I=0$ and $z=+1$.  In order to determine them we impose the 
following conditions: 

\noindent
{\bf 1.} $F_C(s)-B_C(s)$ vanishes linearly in $s$ for $s\to 0$ and we match the
 coefficient to the one loop $\chi$PT result \cite{bc88,dhl87}. Here $B_C$ is the Bron term 
 for $\gamma\gamma\to \pi^+\pi^-$.

\noindent  
{\bf 2.} $F_N(s)$  vanishes linearly for $s\to 0$  and the coefficient can be 
 obtained again from  one loop $\chi$PT \cite{bc88,dhl87}.

\begin{figure}
\includegraphics[width=0.5\textwidth,angle=0]{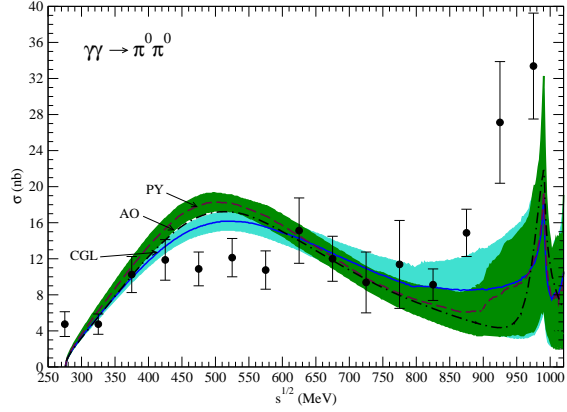}
\caption{Final result for the $\gamma\gamma\to\pi^0\pi^0$ cross for $\sqrt{s}\leq 1.05$~GeV.
 The experimental data are from the Crystal Ball Collaboration \cite{crystal}. 
 }
\label{fig:result}
\end{figure}

\noindent
{\bf 3.} For $I=0$ and $z=+1$ one has in addition the constant $F_0(s_1)-L_0(s_1)$. Its value 
 can be restricted because  the  cross section
  $\sigma(\gamma\gamma\to\pi^0\pi^0)$ around the $f_0(980)$ resonance is quite sensitive
  to this constant. We impose that
   $\sigma(\gamma\gamma\to \pi^0\pi^0)\leq 40$~nb at $s_1$. 
This upper bound for the peak of the $f_0(980)$ 
in $\gamma\gamma\to\pi^0\pi^0$ is equivalent to impose that the $\gamma\gamma$ width of the 
$f_0(980)$ lies in the range
$205^{+95}_{-83}(stat)^{+147}_{-117}(sys)$~eV as determined in ref.\cite{mori}. 
 We shall see that the effect of this rather large uncertainty
 allowed at 1~GeV, see fig.\ref{fig:result}, is very mild 
at lower energies. As the $f_0(980)$ resonance gives rise to a small {\it peak} in the precise data 
 on $\gamma\gamma\to\pi^+\pi^-$ \cite{mori}, then $\phi_0(s)$ must increase with energy above $s_K$
and the case with $z=+1$ is the one realized in nature. 
 Note that for $z=-1$ in eq.(\ref{f0f}), 
there is no a local maximum associated with
 this resonance in $|F_0(s)|$ but a minimum, because $|\omega_0(s)|$ has a dip 
around the $f_0(980)$ mass. 

The source of uncertainty in the approximate relation $\phi_0(s)\simeq \delta_\pi(s)_0$ 
for  $4m_K^2\lesssim \sqrt{s}\lesssim 1.5$~GeV and its functional dependence for $s>s_H=2.25$~GeV$^2$
 is estimated similarly as 
 in ref.\cite{orlast,or07}.
 In fig.\ref{fig:result} we show our final results for the $\gamma\gamma\to \pi^0\pi^0$, where the 
 band around each line corresponds to the estimated error. The
  error band for the dot-dashed line is not shown because it
   is similar to the ones of the other two curves.  In this figure
PY refers to using the $I=0$ S-wave $\pi\pi$ of ref.\cite{py03}, CGL that of ref.\cite{cgl} and 
AO the one of ref.\cite{alba}. One observes that for $\sqrt{s}\lesssim 0.8$ GeV the uncertainty in the
loose bound for the $f_0(980)$ greatly disappears. For such energies the main source of uncertainty
originates from the uncertainties in the $\pi\pi$ phase parameterizations used.
 
\subsection{The $\sigma\to\gamma\gamma$ width}

For the evaluation of the coupling  $\sigma\to \gamma\gamma$, $g_{\sigma\gamma\gamma}$, 
 the amplitude $F_N(s)$ has to be evaluated on the second Riemann sheet, 
 with  $q_\pi\to -q_\pi$.  We denote by $\widetilde{F}_0(s)$ and $T_{II}(s)$ the $I=0$ 
 S-wave amplitudes 
 evaluated on this sheet.  Both have a pole corresponding to the $\sigma$ resonance at 
$s_\sigma$, so that 
\be
T_{II}^{I=0}=-\frac{g_{\sigma \pi\pi}^2}{s_\sigma-s}~,~\widetilde{F}_0(s)=\sqrt{2}\,\frac{g_{\sigma
\gamma\gamma}g_{\sigma\pi\pi}}{s_\sigma-s}~,
\label{coupres}
\ee
where  $g_{\sigma \pi\pi}$ is 
the  coupling of the $\sigma$ to $\pi\pi$.
  The relation between $F_0$ and
$\widetilde{F}_0$ can be easily
established using unitarity above the $\pi\pi$ threshold \cite{orsletter,orlast} so that
$
\widetilde{F}_0(s)=F_0(s)\left(1+2i \rho(s)T_{II}^{I=0}(s)\right).$ 
Taking this into account together with eq.(\ref{coupres})  it follows that
\be
\frac{g_{\sigma\gamma\gamma}^2}{g_{\sigma\pi\pi}^2}=-\frac{1}{2}\left(\frac{\sigma_\pi(s_\sigma)}{8\pi}\right)^2 
F_0(s_\sigma)^2~.
\label{cdespejado}\ee

We denote by $s_\sigma=(M_\sigma-i\,\Gamma_\sigma/2)^2$. Ref.\cite{caprini} provides 
  $M_\sigma^{CCL}=441^{+16}_{-8}~$MeV and 
$\Gamma_\sigma^{CCL}=544^{+18}_{-25}~$MeV, while from ref.\cite{alba} one 
has $M^{AO}_\sigma=(456\pm 6)$~MeV and $\Gamma^{AO}_\sigma=(482\pm 20)$~MeV. 
 In the following the superscripts $AO$ and $CCL$ refer to 
those results obtained by employing $s_\sigma$ from ref.\cite{alba} or \cite{caprini}, respectively. 
   From eq.(\ref{cdespejado}) we obtain 
    $|g_{\sigma\gamma\gamma}/g_{\sigma\gamma\gamma}|=2.01\pm 0.11$ for $s^{CCL}_\sigma$ 
    and $1.85\pm 0.09$ for $s^{AO}_\sigma$.
 Given $s_\sigma$, this ratio of residua is the well defined prediction
  that follow from our $F_0(s)$. 
 We employ the standard narrow resonance width formula in terms of
  $g_{\sigma\gamma\gamma}$ to calculate  
  $\Gamma(\sigma\to \gamma\gamma)=\frac{|g_{\sigma\gamma\gamma}|^2}{16\pi M_\sigma}~$.
 One needs to provide numbers for $|g_{\sigma\pi\pi}|$ in order to apply the previous equation  
  and the determined $|g_{\sigma\gamma\gamma}/g_{\sigma\pi\pi}|$. 
 We first consider the value  $|g_{\sigma\pi\pi}^{AO}|=(3.17\pm 0.10)$~GeV from the 
 approach of ref.\cite{alba}.  The calculated width is  
 $\Gamma^{AO}(\sigma\to\pi\pi)=(1.50\pm 0.18)~\hbox{KeV}.$
 Not only the position of the pole in the partial wave amplitude, but also its residue 
can be calculated in the framework of the dispersive analysis described in ref.\cite{caprini}. 
Expressed in terms of the complex coefficient $g_{\sigma \pi \pi}$ defined 
in eq.(\ref{coupres}), the preliminary result for the residue amounts 
to $|g^{CCL}_{\sigma \pi \pi}| =(3.31^{+0.17}_{-0.08})$~GeV, 
$\Gamma^{CCL}(\sigma\to\gamma\gamma)=(1.98^{+0.30}_{-0.24})~\hbox{KeV}.$
 Taking the average between these two values for $\Gamma(\sigma\to\gamma\gamma)$  we end with,
\be
\Gamma(\sigma\to \gamma\gamma)=(1.68\pm 0.15)~\hbox{KeV}~.
\label{gammafinal}
\ee

\section{On the ratio of couplings $f_0(980)\to \pi\pi/f_0(980)\to K\bar{K}$}

In ref.\cite{npa} the $I=0$ and 1 S-wave meson-meson amplitudes were studied
 using lowest order
Chiral Perturbation Theory (CHPT) to provide the interaction kernels which 
were then implemented in a Bethe-Salpeter equation. With only one free parameter
 the resonances $\sigma$, $f_0(980)$ and $a_0(980)$ were generated and 
 the scattering data reproduced properly.
This paper was the basis for all the later developments in Unitary CHPT.
 According to this approach, the $I=0$ S-wave with  the
 $\pi\pi$ and $K\bar{K}$ channels can be written in matrix notation as:
 \be
  T=\left[I+V\cdot G\right]^{-1}\cdot V~,~V=\left(\begin{matrix} V_{11} & V_{12}\\ V_{12} & V_{22}\end{matrix}\right)
~,~G=\left(\begin{matrix} g_1 & 0\\ 0& g_2\end{matrix}\right)~,
\label{mat}
 \ee
 with $V_{11}=(s-m_\pi^2/2)/f_\pi^2$, $V_{12}=\sqrt{3}s/4f_\pi^2$,
  $V_{22}=3s/4f_\pi^2$ and the labels 1 and 2  
   refer to the $I=0$ $\pi\pi$ and $K\bar{K}$ states, in order. The function
   $g_i(s)$ is given by
   \be
   g_i(s)=\frac{1}{16\pi^2}\left(\alpha_i+\log
   \frac{m_i^2}{\mu^2}-\sigma_i(s)\log\frac{\sigma_i(s)-1}{\sigma_i(s)+1}\right)~,
\label{gs}   \ee
where $\sigma_i(s)=\sqrt{1-4m_i^2/s}$. The constant $\alpha_i$ can be fixed by comparing the previous
  expression 
 with the one calculated in terms of a three-momentum cut-off, the latter 
can be found in ref.\cite{prdiam}. One has then
 $\alpha_i=-\log(1+\sqrt{1+m_i^2/\Lambda_\chi^2})^2+\log \mu^2/\Lambda_\chi$, 
 with $\Lambda_\chi\simeq 1~$GeV, the scale of the CHPT suppression parameter. 
 On the other hand, $m_1$ and $m_2$ are the pion and kaon masses, respectively.  
Taking the ratio of  $T_{12}$ and $T_{22}$ from 
eq.(\ref{mat}) one has:
\be
\frac{T_{12}}{T_{22}}=\frac{1/\sqrt{3}}{1+g_1 3s/4f^2}~.
\label{rat1}\ee
Since $s\simeq 1$~GeV$^2$ we have neglected the factor $2 m_1^2\ll 3s$ in 
$3s-2m_1^2$. 
When inverting the matrix $[I+V\cdot G]$ in eq.(\ref{mat}) the zeroes of its
determinant determine the positions of the resonance poles. This determinant
is given by $1 + V_{22} g_2 +  V_{11} g_1+ (V_{11}V_{22}-V_{12}^2)g_1
g_2$ from where at the pole position, $s_R$, one has 
 $V_{22}(s_R)=-\frac{g_1}{g_2}\left[1-V_{12}^2 g_1 g_2/(1+V_{11}g_1)\right]$.
 This expression can be substituted in eq.(\ref{rat1}) because
 $3s/4f^2=V_{22}$, with the result,
 \be
\frac{T_{12}}{T_{22}}=\frac{1/\sqrt{3}}{1 -
\frac{g_1}{g_2}\left(1-V_{12}^2 g_1 g_2/(1+V_{11}g_1))\right)}~.
\label{rat2}
 \ee
 Now eq.(\ref{mat}) for $T$ is equivalent to $
T=V+V\cdot G\cdot T~.$ 
 As $V\propto f^{-2}\propto N^{-1}_c$ and $g_i={\cal O}(N_c^0)$, 
 with $N_c$ the number of colours, it is necessary
 that the pole positions of  the $\sigma$, $f_0(980)$ (and also for the $a_0(980)$) run
 as $f^2$.  In this way, $V_{ij}(s_R)\propto s_R/f^2={\cal O}(N_c^0)$ 
  and there is no 
  a mismatch in the way that $T$ runs with $N_c$ at $s_R$ 
  between the left and right sides of the previous expression.
 The same result was obtained in ref.\cite{nd} considering directly 
 the dependence on $N_c$ of the solutions of the equation $1+V\cdot G=0$.  Thus,  eq.(\ref{rat2})
is ${\cal O}(N_c^{0})$. It is worth stressing that this equation
 corresponds to the ratio  of the $f_0(980)$ couplings to
$\pi\pi$ and $K\bar{K}$, $ \gamma_\pi$ and $\gamma_K$, respectively, when 
 evaluated at the $f_0(980)$ pole
position. As a result
this ratio does not run with $N_c$ and the suppression of the $\pi\pi$ coupling
compared to the $K\bar{K}$ one, around a factor 3 smaller as given by 
eq.(\ref{rat2}), does not stem from the OZI rule.  Let us recall that within QCD
 the OZI rule is a requirement of the large $N_c$ limit.
This suppression of the $\pi\pi$ coupling originates from the
factor $1/\sqrt{3}$ in eq.(\ref{rat2}), required by the expressions of $V_{ij}$ 
from lowest order CHPT, and from the rescattering, accounted for by the denominator in 
eq.(\ref{rat2}). 

In summary, we have presented a brief account of the results of refs.\cite{orsletter,orlast} on the 
calculation of $\gamma\gamma\to\pi\pi$ from dispersion relations and of the width 
$\Gamma(\sigma\to\gamma\gamma)$. In addition, we have shown that the ratio of the $f_0(980)$ couplings to $\pi\pi$ and
 $K\bar{K}$ does not run with $N_c$, so that the suppression of the former compared with the latter
 has nothing to do with the OZI rule. 

\begin{theacknowledgments}
  Financial support from the grants MEC  FPA2007-6277 and
      Fundaci\'on  S\'eneca  02975/PI/05 is acknowledged. 
\end{theacknowledgments}



\end{document}